# High-dimensional frequency crystals and quantum walks in electro-optic microcombs

Christian Reimer[1,2,*], Yaowen Hu[1,3,*], Amirhassan Shams-Ansari[1],
Mian Zhang[1,2], Marko Loncar[1,+]

[1]John A. Paulson School of Engineering and Applied Sciences, Harvard University, Cambridge, MA 02138, USA
[2]HyperLight Corporation, 501 Massachusetts Ave, Cambridge, MA 02139, USA
[3]Department of Physics, Harvard University, Cambridge, MA 02138, USA
*These authors contributed equally
+email: loncar@seas.harvard.edu

**Synthetic optical structures[1] have enabled studies of various physical processes and discoveries of new phenomena in a controllable manner[2–7]. Past realizations, using for example coupled optical waveguides[1–9], were however limited in size and dimensionality. Here we theoretically propose and experimentally demonstrate optical frequency crystals of arbitrary dimensions, formed by hundreds of coupled spectral modes within an on-chip electro-optic frequency comb[10]. We show a direct link between the measured optical transmission spectrum and the density-of-state of frequency crystals, validated by measurements of one-, two-, three- and four-dimensional crystals with no restrictions to further expand the dimensionality. We measure signatures of quantum random walks and Bloch oscillations in the frequency domain of light in one- and two-dimensional synthetic lattices. Our work employs a quantum description of phase modulation[11] to describe resonant electro-optic frequency combs, and shows that they enable the study of exotic crystal properties, topologies and quantum dynamics, while providing a versatile platform to manipulate quantum information in the optical frequency domain.**

While nearly all solid-state crystal structures occurring in nature are three-dimensional, recent research efforts have resulted in the discoveries of lower-dimensional structures such as graphene or carbon nanotubes. Increasing the dimensionality of solid-state crystals is however implausible since they are bound by the three-dimensional Euclidian space. Extending the concept of high-dimensional spaces to crystal structures is an intriguing concept, which has been studied from a theoretical standpoint[12,13]. In contrast to solid-state crystals, synthetic structures, realized by e.g. optical lattices[1–9], have recently enabled the investigation of various physical processes in a controllable manner[2–7], and even the study of new phenomena[8,9]. High-dimensional synthetic crystal structures are of significant interest: they can be used to investigate complex dynamics of solid-state materials, where e.g. the impact of forces, gauge fields, defects or multi-particle interactions could be mapped to higher dimensions[14–16]. Furthermore, by mapping one system onto another with higher dimensionality, it becomes possible to solve certain problems more efficiently, which is the working principle of reservoir computers[17]. Synthetic crystals are also ideally suited to study complex dynamics in a highly controllable manner, since they are not restricted by physical space and can thus provide unique properties including high-dimensionality. Optics, in particular, provides a powerful platform since the modes of light can be described by the same equations that govern the dynamics of many other physical systems. The realizations of synthetic optical structures have included measurements of classical and quantum correlations[2], Bloch oscillations[3], Anderson localization[4], Ising spin chains[5], quantum random walks[6], topological structures[7], as well as parity-time[8] and super-symmetric[9] lattices. Past realizations of synthetic photonic lattices were, however, limited in size and dimensionality, as they mainly relied on coupled optical waveguides[3–9], i.e. a spatial degree of freedom on a two-dimensional plane.

Here we investigate the quantum properties of electro-optic frequency combs[10] and show that they can be modeled by means of a generalized tight-binding model. In agreement with the theoretical predictions, we measure the generation of synthetic optical frequency crystals in high-dimensional space. In our approach, discrete lattice points are formed by spectral modes of an optical microring resonator realized in thin-film lithium niobite (LN) integrated photonic platform[18], while the coupling between lattice points is controlled by electro-optic phase modulation[11], enabled by the second-order nonlinearity of LN. Light coupled into the frequency crystals experiences quantum coherent scattering and interference at different lattice points, in direct analogy to electron behaviour in solid-state crystals. We show that it is possible to directly measure the density-of-states (DOS) of the frequency crystals and furthermore measure signatures of quantum coherent scattering processes such as Bloch oscillations and multi-dimensional quantum random walks. Our results show that electro-optic frequency combs can be used for frequency-domain quantum information processing[19–23].

The tight-binding model is one of the most fundamental models in solid state physics[24], which assumes that particles (such as electrons or here photons) are localized at specific positions of the crystal lattice, and that they can hop between neighboring lattice points while preserving phase coherence. Optical tight-binding systems have in the past been realized using spatial modes in coupled optical waveguides[2–4,25], or temporal modes in coupled resonators with different roundtrip times[8,26]. In contrast, here we experimentally realize synthetic crystal[15,27–29] utilizing the discrete frequency modes of a LN microring resonator[10], see Fig. 1 and Methods. By applying an electronic radio-frequency (RF) signal with a frequency equal to the separation between adjacent frequency modes (known as free spectral range, FSR), optical coupling can be initiated, where the coupling strength can be adjusted by the strength of the electric driving signal. Such an electro-optic resonator driven by a single-tone RF signal, commonly referred to as an electro-optical frequency comb[10,30], can also be described in a quantum model as an one-dimensional tight-binding lattice[28] with a hopping rate related to the applied RF power. We here show that such a tight-binding frequency crystal representation is not limited to one-dimensional realizations. For example, using two RF signals (both only very slightly detuned from the resonator FSR), a photon placed in one optical resonance (Fig. 1b) can be frequency-shifted to neighboring resonance by the driving RF signal. Such a system can be described by a two-dimensional tight-binding model, see Methods. Remarkably, we demonstrate that the same principle can be extended to three, four and many more dimensions using different RF driving signal. Here, each additional RF frequency-tone can span an additional spectral dimension, see Fig. 1b and Methods. Importantly, individual frequency modes within the microring resonators can be unambiguously mapped to individual lattice points within the crystal's synthetic frequency space, see Fig. 1. This high level of control and one-to-one mapping of spectral modes in real frequency space to lattice points in synthetic frequency space enables the experimental investigation of crystal structures in high-dimensions.

Using the quantum mechanical tight-binding model to describe the electro-optic frequency comb, we show that the frequency crystal's density-of-states (DOS) can be directly measured by probing the optical transmission spectrum of the resonator, see Methods. In particular, using input-output theory of optical resonators, we established a direct relationship between the optical transmission $T(\Delta)$ and density of states $D(\Delta)$ of the frequency crystals, see Methods, which is given by:

$$T(\Delta) = 1 - 2\pi \frac{\kappa_e}{N_t^d} D(\Delta)$$

where Δ is wavelength detuning from the resonance center, $\kappa_e$ is the external coupling rate, $d$ is the number of RF tone, and $N_t$ is the total number of cavity resonance mode coupled by one RF tone, see Methods. The expression can be understood by considering that a larger DOS leads to a the larger the number of optical modes that are coupled through electro-optic modulations, and due to the interference condition in the resonator, this leads to a larger the effective optical "absorption" within the resonator, and thus smaller transmission. These results show that measuring the transmission spectrum of cavity resonance, using a tunable continuous waver laser at telecom wavelengths, represents a direct measurement of the DOS of the frequency crystal. Leveraging this analogy, we experimentally probed the DOS of one-, two-, three-, and four-dimensional frequency crystals, using up to four RF drives to drive an electro-optic ring resonator with 10 GHz free spectral range, where the RF tones were detuned by 1 MHz with respect to each other. We found excellent agreement with the analytical solution for the DOS of the crystals, see Fig. 2.

A particularly important feature described by a tight-binding model is the occurrence of quantum random walks, which arise from the phase-coherent step-wise propagation of particles in a lattice. Therefore, exciting our frequency crystal with a photon that has a narrow spectral linewidth (i.e. driving a single lattice point of the crystal in synthetic space) is expected to give rise to the quantum random walk dynamics in frequency domain, resulting in spectral spreading for each roundtrip in the resonator (Fig. 3a). However, a photon that is spectrally narrow enough to only excite a single resonance intrinsically has to have a temporal duration much longer than the round-trip time of the resonator. For this reason, a description using individual round-trips cannot adequately describe the dynamics. We find that instead of experiencing discrete steps, multiple steps of the quantum walk coherently interfere over the coherence time of the photon, forming a steady-state output, see Methods and Fig. 3c. This result of our tight-binding model, predicting an exponential envelope in the output spectrum, agrees with classical model of electro-optic frequency combs[10], and therefore provides a means to model electro-optical combs at the single photon level.

Bloch oscillations are another well-known effect in solid-state physics, which occur in the presence of a linear force in the crystal. Theoretical considerations of one-dimensional frequency crystals have predicted that a linear force can be induced if the RF driving field is detuned from the spectral separation of the microring resonator modes[28]. In particular, with an RF modulation frequency significantly detuned from the FSR, spectral modes are generated detuned from the center of the resonances, which in turn induces additional phase-shift. This effect is analogous to a phase shift induced by a linear force in solid-state crystals that are responsible for Bloch oscillations[28]. In the frequency crystal, these Bloch oscillations result in the re-localization of the light at the input frequency after a certain number of roundtrips, see Fig. 3b. When excited with a spectrally narrow photon, multiple roundtrips coherently interfere over the coherence time of the photon, and the resulting steady state solution shows a spectrum with interference fringes and clear cut-offs in the spectrum, which arises from the Bloch oscillations, see Fig. 3d. The measured optical spectrum agrees well with simulations and represents a measured signature of Bloch oscillations in the frequency domain of light.

Considering two- and higher-dimensional frequency crystals, quantum random walks are expected to appear in the synthetic frequency space (schematic in Fig. 1b), which maps into measurable signatures in real frequency space (schematic in Fig. 1c). In a two-dimensional frequency crystal

(formed by two RF modulation frequencies), a photon positioned in a single lattice point can "hop" to one of the four nearest-neighbour lattice points per roundtrip. Fig. 4a shows examples of possible paths that a photon can take if placed at a single lattice point (e.g. at the center of Fig. 4a) within a 2D synthetic lattice. In classical random walks, the different possible paths are independent, leading to a Gaussian probability distribution. In contrast, in quantum random walks, all possible paths to individual lattice points are phase coherent and interfere, leading to a non-Gaussian probability distribution. Making use of the tight-binding model, we numerically simulated the probability distribution of a single photon propagating in a frequency crystal, finding clear signatures of quantum random walks, see Fig. 4b. Furthermore, as discussed in Fig. 3, when frequency crystal is excited with spectrally narrow photons, multiple random-walk steps coherently interfere, leading to a steady-state photon distribution in synthetic frequency space, see Fig. 4c. This steady-state is dominated by the shape of the quantum random walks for different roundtrips and shows a clear localization of the photons to a narrow spectral bandwidth far away from the excitation wavelength. It is possible to map the photon distribution in synthetic space to the measurable real frequency space (see Fig. 1c), where the spectral content within resonances of the microring resonator represent cross sections through the synthetic frequency space, see Fig. 4c. Our theoretical considerations predict that the signatures of two-dimensional quantum random walks lead to many excited spectral modes within resonances close to the excitation frequency, and to only very few frequency modes for resonances far away from the excitation. This is confirmed experimentally using a heterodyne detection technique to measure the spectral content of the resonances close to and far away from the excitation one, see Fig. 4e-h. In particular, for cavity resonances close to the excitation frequency we measure many excited spectral modes (see Fig. 4e-f), while for cavity resonances further away from the excitation, the measured spectral content narrowed down and consisted only of a few lines (see Fig. 4g-h). Our measurements confirm the existence of spectral narrowing, a key signature of two-dimensional quantum random walks. This spectral localization is highly relevant for the generation of broad EO combs. Indeed, it has been shown that the spectral extent of EO combs is mainly limited by a cut-off imposed by detuning of spectral modes from them center of the cavity resonances[10]. Remarkably, the quantum random walks in frequency crystals counters spectral cut-offs and eliminates the effect of Bloch oscillations in the two-dimensional quantum walk. In particular, the quantum walks force the photons to propagate mainly within the center of the resonances where it does not experience resonator-induced phase shifts (i.e. linear forces), therefore enabling broad comb generation when driven with multiple RF tones.

In conclusion, we have theoretically proposed and experimentally generated high-dimensional synthetic frequency crystals by driving an electro-optic resonator with multiple RF signals. Optical transmission measurements are demonstrated to be a powerful tool for the direct characterization of the density-of-state of frequency crystals. A generalized tight-binding formalism has been applied to model the quantum properties of resonant electro-optic frequency combs, which was able to predict the measured signatures of Bloch oscillations and quantum random walks in one- and two-dimensional frequency crystals. Furthermore, we performed the first experimental characterizations of tight-binding systems with dimensionality larger than three. Our work provides fundamental insight into the nature of electro-optical frequency combs, where the tight-binding description of the system provides a new understanding of the comb dynamics, especially when driven with multiple RF signals, and enables modeling of the system of single photons. While the performed measurements themselves were classical in this work, the measured signatures nevertheless confirm that the photon dynamics within the frequency crystals are quantum coherent and phase stable. Future work will focus on investigating two- or multi-photon dynamics, which can be described in the here-presented theoretical framework. This opens up new possibilities of using electro-optical combs for frequency domain quantum information processing[23]. In particular, non-resonant electro-optic modulation has been used for the implementation of quantum state manipulation in the frequency domain[31], and the excitation of frequency crystals with squeezed[19] or frequency-entangled[31] optical quantum states could be used to perform various quantum operations. Electro-optic frequency combs are therefore a versatile platform to investigate synthetic crystal structures, and furthermore provide a powerful tool to perform quantum information processing in the frequency domain of light.

**Acknowledgements.** We thank J. MacArthur for providing an RF source for the experiments, Cheng Wang and Shunyu Yao for helpful discussions, and B. Machielse and S. Bogdanovic for feedback on the manuscript. This work is supported in part by the National Science Foundation (NSF) (ECCS1609549, ECCS- 1740296 E2CDA and DMR-1231319) and by Harvard University Office of Technology Development (Physical Sciences and Engineering Accelerator Award). Device fabrication was performed at the Harvard University Center for Nanoscale Systems, a member of the National Nanotechnology Coordinated Infrastructure Network, which is supported by the NSF under ECCS award no. 1541959.

**Figures:**

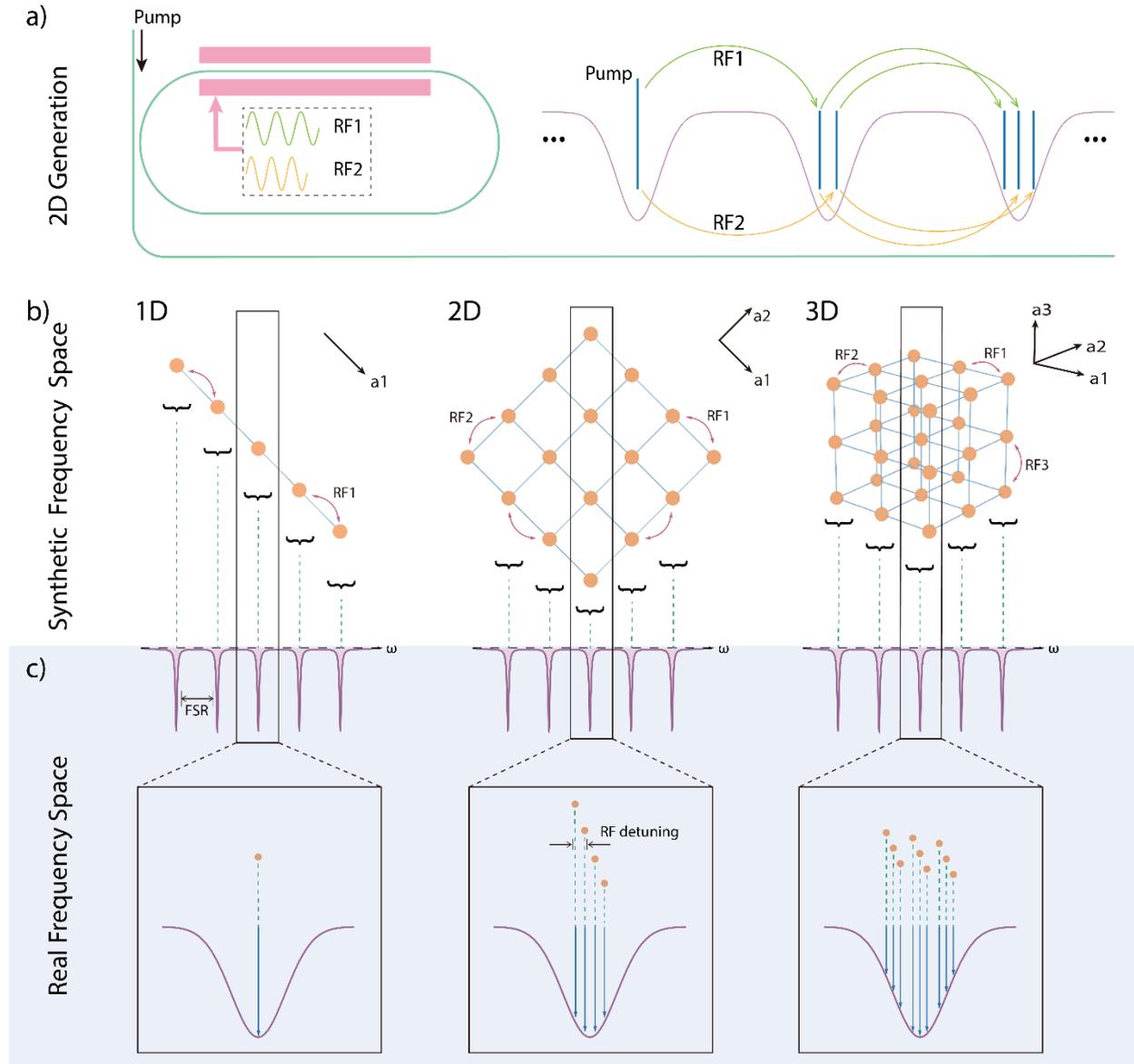

**Fig. 1: Optical frequency crystals generation in electro-optic frequency combs. a)** Schematic of the electro-optically modulated resonator used to generate the frequency crystals. The device consists of a waveguide coupled race-track resonator with electrodes placed around it. The high-dimensional frequency crystals are formed by modulating the device with multiple, slightly detuned, RF signals (here two RF signals are shown for illustration). This gives rise to multiple excitations of each race-track resonance, each representing a crystal lattice point. **b)** The tight-binding crystals can be represented in synthetic frequency space as fixed lattice points (yellow circle), where coupling between neighboring lattice points is mediated by electro-optic modulation due to applied RF field. By modulating simultaneously with different frequencies, high-dimensional lattices can be generated in synthetic frequency space. **c)** As a result, each optical resonance (mode of a resonator) represents one lattice point, one crystallographic direction, or one crystal plane of the synthetic frequency crystal when one, two or three RF tones are applied,

respectively. Within each resonance, the spacing between optical excitations is determined by the frequency difference between the RF driving signals.

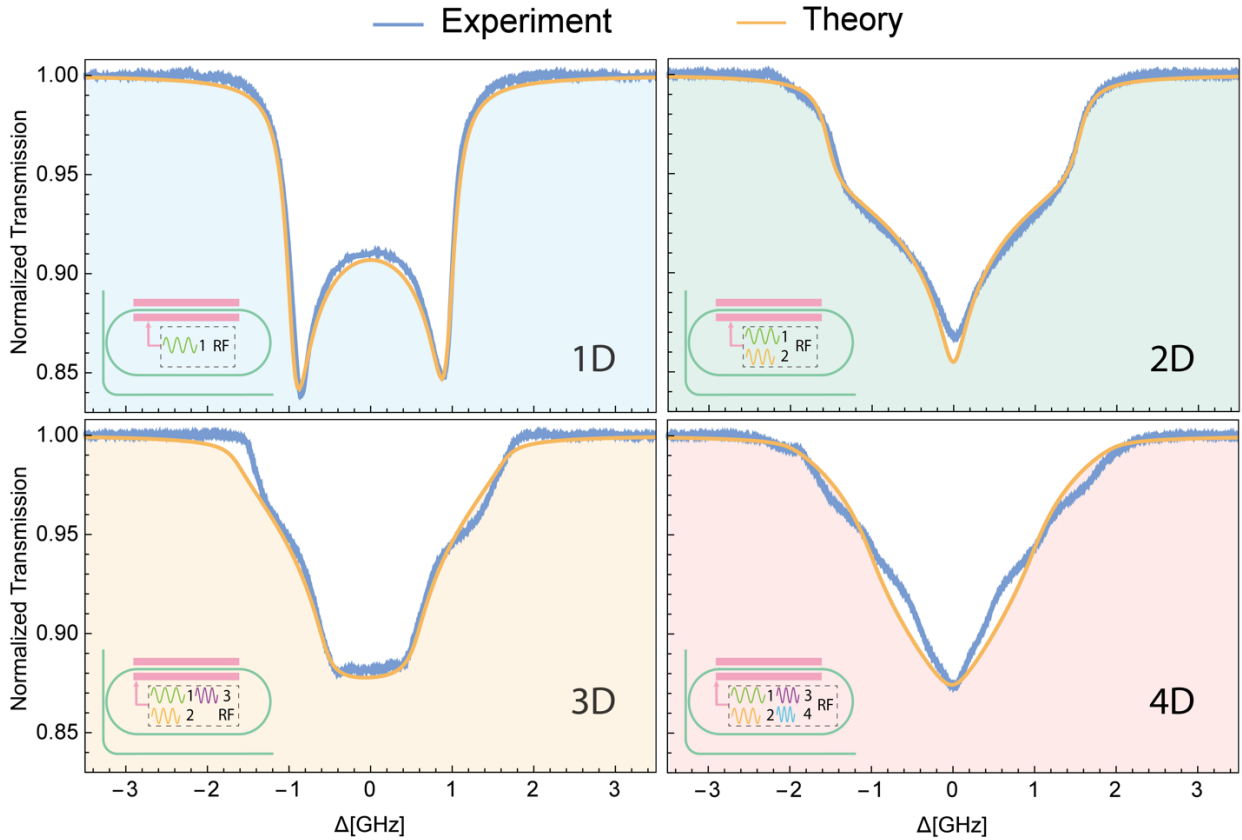

**Fig. 2: Density-of-states of high-dimensional frequency crystals.** By scanning an excitation laser through the optical resonances, it becomes possible to directly measure the density-of-states of the frequency crystals. Shown is the measured normalized optical transmission (blue trace) for one- to four-dimensional crystals, superimposed with the analytical model (orange trace) based on the density-of-states of the frequency crystal. Very good agreement is observed, thus confirming that the density-of-states can be measured directly.

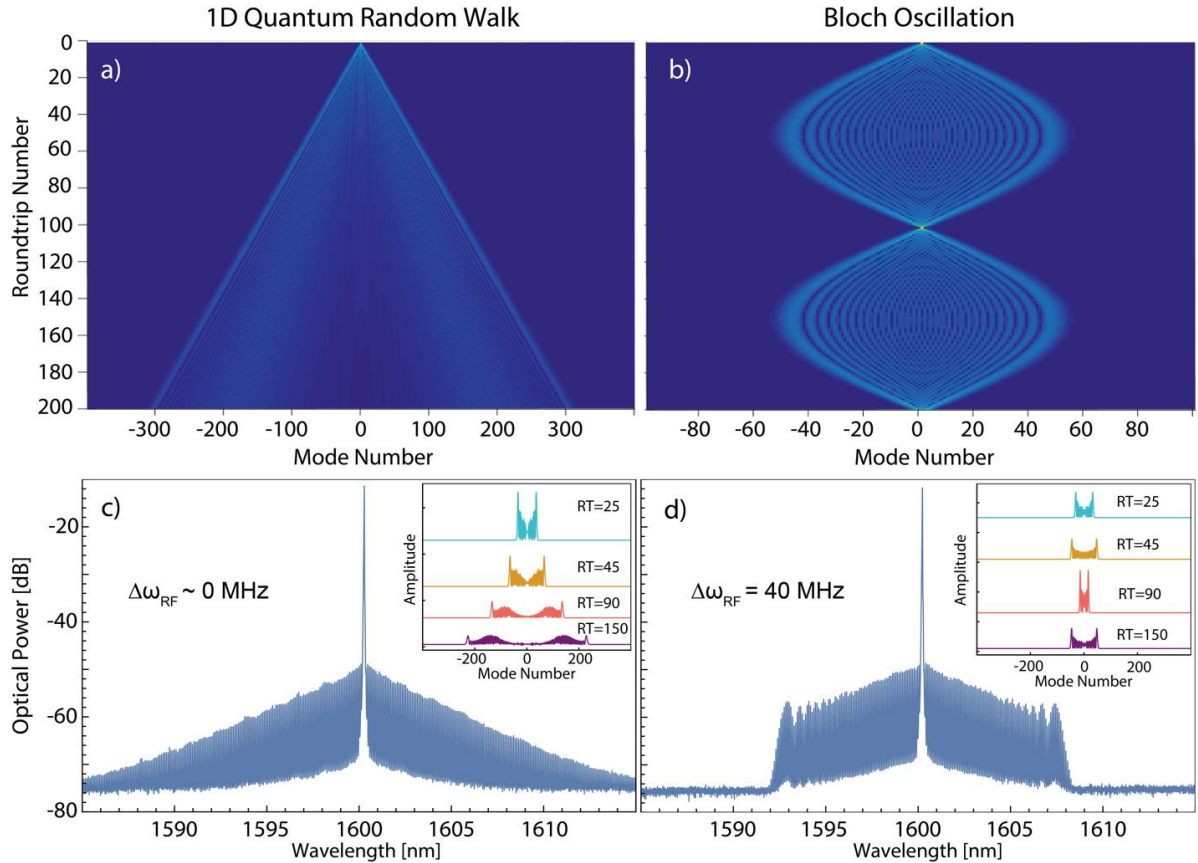

**Fig. 3: Quantum random walks and Bloch oscillations in one-dimensional frequency crystals.**
**a)** In the absence of RF fields, CW laser excitation of the ring resonator leads to the excitation of one spectral mode (mode number 0), which is equivalent to the excitation of a single lattice point of a synthetic crystal. With RF fields applied, photon can hop to neighboring optical modes, giving rise to a quantum random walk and spectral broadening. Numerical simulations show that spectral modes with increasing mode number can be excited as light completes more roundtrips (RT) inside the resonator. Here, the frequency of the RF drive was perfectly matched to the free spectral range (FSR) of the resonator. **b)** If the RF driving signal is detuned form the resonator FSR, an effective linear force is imposed, which leads to Bloch oscillations in the frequency domain. **c)** If the frequency crystal is excited with photons that are spectrally narrow enough to excite only a single lattice point, their temporally duration has to be larger than multiple round trips. In such narrow-band excitation, all roundtrips of the random walks coherently interfere over the coherence time of the photon, forming a steady state output with characteristic exponentially decaying spectrum. **d)** In the presence of Bloch oscillations, a sharp cut-off in the optical output spectrum is measured, which arises from the oscillations in the random walks. The insets in **c)** and **d)** show numerical simulations for different RTs to illustrate the effect arising from the coherent addition of multiple quantum random walk roundtrips.

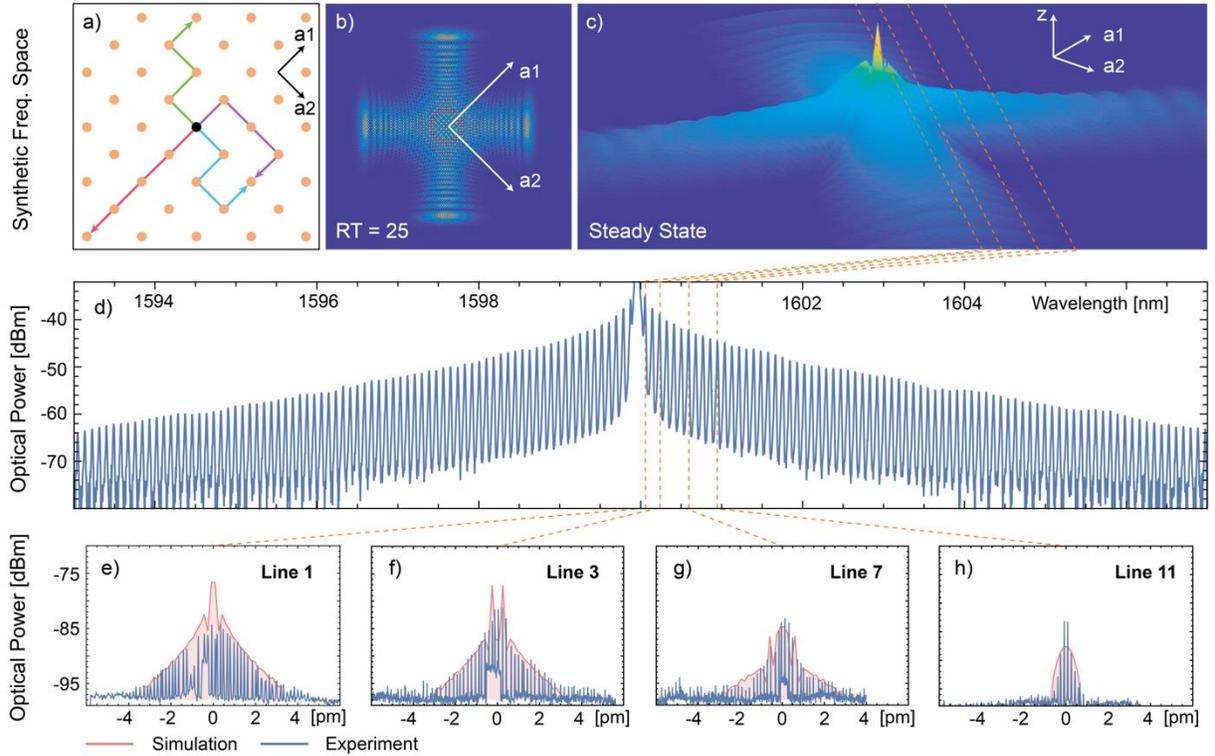

**Fig. 4: Probing quantum walks in two-dimensional synthetic crystals. a)**. Illustration of possible paths that photon can take after four random walk steps (i.e. round trips), if placed in the center of the 2D synthetic lattice. If the photon propagation is quantum coherent, all possible paths leading to the same lattice point interfere, resulting in interference patterns in the probability distribution of the photon location in the lattice, see **b)** for 25 simulated roundtrips (RT). If the frequency crystal is excited spectrally narrow photons (single lattice point excitation), multiple roundtrips (RT = 1, 2, …, N) coherently interfere, leading to a steady-state photon probability distribution in synthetic frequency space (the color and z-axis represents the amplitude), see **c)**. The distribution in synthetic space can be mapped to measurable real frequency space, see Fig. 1. The emission spectrum in real frequency space was measured with an optical spectrum analyzer **d)**, while the spectral content within individual resonances was measured using homodyne detection **e-h)**. As the spectral content within individual resonances are traces through the steady-state in synthetic space, broad spectra were expected and measured close to the excitation **e-f)**, while narrow spectra were measured further away from the excitation, **g-h)**, showing excellent agreement with simulations.

## Methods:

**Experimental setup.** The electro-optic resonators were fabricated on x-cut single crystalline 600 nm thick lithium niobate (LN) layer on top of 2 µm thermally gown $SiO_2$ layer, on top of silicon substrate (wafers by NANOLN). The patterns were defined using standard electro-beam lithography, and were transferred into the LN layer using argon plasma etching. Gold electrodes were placed next to the optical waveguides to enable electro-optic modulation. The chips were diced and the facet polished to enable end-fire optical coupling implemented with two lensed optical fibers, and the electrical signal was fed to the microelectrodes using a contact probe. The microwave driving signals were generated by four radio-frequency (RF) synthesizers (two HP 83711B and two Hittite HMC-T2240), which were phase-locked via a common 10 MHz clock and can operate independently at different frequencies. The different RF signals were combined using RF power splitters and were amplifier by an electrical power amplifier (Pasternack) and passed through a circulator with a 50 Ohm terminated third port, before being connected with the contact probe. In all presented measurements, we used a tunable laser (Santec TSL-510) to pump the resonator. For the measurements shown in Fig, 2 the laser was scanned through the resonances and detected the output with a 125 MHz photodiode (New Focus 1811). For the measurements shown in Fig. 3, the excitation laser was into the center of the resonance and the output spectrum was measured with a standard grating based optical spectrum analyzer with 20 pm resolution. For the heterodyne detection measurements shown in Fig. 4, the excitation laser was tuned to the center of the resonance as for the measurements in Fig. 3, and the light was sent to a 3dB fiber direction coupler and overlapped with light emitted from a second tunable laser (Santec TSL-510), used as a local oscillator, before being detected with a 12 GHz photodiode (New Focus 1544A). The photodiode signal was analyzed with a radio frequency spectrum analyzer to record the beat-note of the local oscillator with the individual lines emitted from the electro-optic resonator.

**Generalized tight-binding model of electro-optic frequency combs driven by one and multiple radio-frequency signals.**

For *one-dimensional frequency crystal*, one electro-optic modulation $\Omega \cos \omega t$ is applied to the ring resonator. This modulation can induce the hopping between different resonant modes of the cavity, which can be described by the following Hamiltonian:

$$H = \sum_{j=-N}^{N} \omega_j a_j^\dagger a_j + \Omega \cos \omega t \, (a_j^\dagger a_{j+1} + h.c.)$$

Here $\omega_j$ is the frequency of the j-th resonant modes and for now we assume that there is no group velocity dispersion so that all resonant modes have equal spacing with free spectral range (FSR) $\omega_F = \omega_{j+1} - \omega_j$. We choose $\omega_0$ (i.e. j=0) as the mode that is optically excited with a pump laser, and we consider $N_t = 2N + 1$ resonant modes.

Moving into a rotating frame of all resonant modes by doing substitution for each FSR mode $a_j \to a_j e^{-i\omega_j t}$, followed by applying the rotating wave approximation, we obtain

$$H = \sum_j \frac{\Omega}{2} (a_j^\dagger a_{j+1} e^{i\delta t} + h.c.)$$

where $\delta = \omega - \omega_F$. In the following we consider the RF modulation with frequency equal to FSR so that $\delta = 0$ for simplicity.

$$H = \sum_j \frac{\Omega}{2}\left(a_j^\dagger a_{j+1} + h.c.\right)$$

This is equivalent to the one-dimensional tight-binding Hamiltonian in k space

$$H = \sum_k \Omega \cos k \, a_k^\dagger a_k$$

where the mode $a_k$ are defined as:

$$a_k = \frac{1}{\sqrt{N_t}}\sum_j e^{-ikj} a_j \qquad a_j = \frac{1}{\sqrt{N_t}}\sum_k e^{ikj} a_k$$

with k defined as $k = \frac{2\pi}{N_t} l \ (l = -N, \ldots, N)$.

For ***high dimensional frequency crystals***, we consider multiple RF tones to modulate the ring resonator, which means that the RF signals applied on the electrode are described by $\sum_{i=1}^{d} \Omega_i \cos \omega_i t$ with the crystal dimension labeled as *d*, and each $\omega_i$ just has a slight frequency difference (around 1MHz) detuned from the FSR $\omega_F$. Then the Hamiltonian is:

$$H = \sum_{j=-N}^{N}\left(\omega_j a_j^\dagger a_j + \sum_i \Omega_i \cos \omega_i t \left(a_j^\dagger a_{j+1} + h.c.\right)\right)$$

Following the similar procedure for one-dimensional frequency crystal, we move to the rotating frame of each FSR mode and apply the rotating wave approximation:

$$H = \sum_j \sum_i \frac{\Omega_i}{2}\left(a_j^\dagger a_{j+1} e^{i\delta_i t} + h.c.\right)$$

Where $\delta_i = \omega_i - \omega_F$ (i=1,2,3,…,d).

We first consider two-dimensional frequency crystal as an example and then preset the general results for arbitrary dimension. In the case *d = 2* the Hamiltonian is

$$H = \sum_j \frac{\Omega_1}{2}\left(a_j^\dagger a_{j+1} e^{i\delta_1 t} + h.c.\right) + \sum_j \frac{\Omega_2}{2}\left(a_j^\dagger a_{j+1} e^{i\delta_2 t} + h.c.\right)$$

In two-dimensional case, there is more than one mode within a resonant peak of the resonator because light at any given frequency can be modulated to the neighbor resonance with two different choices, related to the different RF frequency. In order to make the mode that represents individual lattice points more clearly, we define the frequency crystal modes that corresponding to each lattice point in a two-dimensional crystal as

$$c_{l_1,l_2} = \frac{1}{\sqrt{N_t}} a_{l_1+l_2} e^{i(l_1\delta_1 + l_2\delta_2)t}$$

Which can also be written as

$$c_l = \frac{1}{\sqrt{N_t}} a_{\Sigma l_i} e^{i l \cdot \delta t}$$

Where $l = (l_1, l_2)$ and $\delta = (\delta_1, \delta_2)$. Here $\mathbf{l}$ represents the two-dimensional lattice position label where $l_1$ or $l_2$ can equal to $-N, -N+1, \ldots, N-1, N$. Since all the other modes generated by the hopping of pump photon, we define the pump mode as $c_{0,0} = \frac{1}{\sqrt{N_t}} a_0$, which is at the center point of the lattice.

In real frequency spectrum if two modes (e.g. $c_{1,1}$ and $c_{2,0}$) are in the same resonant peak, then the summation of their own subscript are same (that means $1+1 = 2+0$) because $c_l \propto a_{\Sigma l_i}$.

It can be seen that $c_l$ satisfies $c^\dagger_{l_1,l_2} c_{l_1+1,l_2} = \frac{1}{N_t} a^\dagger_{l_1+l_2} a_{l_1+l_2+1} e^{i\delta_1 t}$ and $c^\dagger_{l_1,l_2} c_{l_1,l_2+1} = \frac{1}{N_t} a^\dagger_{l_1+l_2} a_{l_1+l_2+1} e^{i\delta_2 t}$.
So for any given j, we have a sets of $(l_1, l_2)$ that can satisfy $l_1 + l_2 = j$ ($l_1$ range from $-N, \ldots, N$ and $l_2 = j - l_1$. All of these corresponded $c_{l_1,l_2}$ have the relation $c^\dagger_{l_1,l_2} c_{l_1+1,l_2} = \frac{1}{N_t} a^\dagger_{l_1+l_2} a_{l_1+l_2+1} e^{i\delta_1 t}$ and $c^\dagger_{l_1,l_2} c_{l_1,l_2+1} = \frac{1}{N_t} a^\dagger_{l_1+l_2} a_{l_1+l_2+1} e^{i\delta_2 t}$. So for any j we can obtain the following relation that

$$a^\dagger_j a_{j+1} e^{i\delta_1 t} + h.c. = \frac{1}{N_t} a^\dagger_{l_1+l_2} a_{l_1+l_2+1} e^{i\delta_1 t} + \frac{1}{N_t} a^\dagger_{l_1+l_2} a_{l_1+l_2+1} e^{i\delta_1 t} + \cdots + \frac{1}{N_t} a^\dagger_{l_1+l_2} a_{l_1+l_2+1} e^{i\delta_1 t}$$

$$= \cdots + c^\dagger_{-1,j-(-1)} c_{-1+1,j-(-1)} + c^\dagger_{0,j-(0)} c_{0+1,j-(0)} + c^\dagger_{1,j-(1)} c_{1+1,j-(1)} + \cdots$$

$$= \sum_{l_1=-N}^{N} (c^\dagger_{l_1,j-l_1} c_{l_1+1,j-l_1} + h.c.)$$

The similar process will also apply to $a^\dagger_j a_{j+1} e^{i\delta_2 t} + h.c.$ and we can obtain the Hamiltonian

$$H = \sum_j \frac{\Omega_1}{2} (a^\dagger_j a_{j+1} e^{i\delta_1 t} + h.c.) + \sum_j \frac{\Omega_2}{2} (a^\dagger_j a_{j+1} e^{i\delta_2 t} + h.c.)$$

$$= \frac{\Omega_1}{2} \sum_j \sum_{l_1=-N}^{N} (c^\dagger_{l_1,j-l_1} c_{l_1+1,j-l_1} + h.c.) + \frac{\Omega_2}{2} \sum_j \sum_{l_1=-N}^{N} (c^\dagger_{l_1,j-l_1} c_{l_1,j-l_1+1} + h.c.)$$

$$= \frac{\Omega_1}{2} \sum_{l_1} \sum_{l_2} (c^\dagger_{l_1,l_2} c_{l_1+1,l_2} + h.c.) + \frac{\Omega_2}{2} \sum_{l_1} \sum_{l_2} (c^\dagger_{l_1,l_2} c_{l_1,l_2+1} + h.c.)$$

Which is the two-dimensional tight-binding model. By expressing the Hamiltonian in mode $c_l$, we are changing the time-dependence of the operator, which will involve some extra term in the Hamiltonian that is proportional to $\delta_1, \delta_2$. But since $\delta_1, \delta_2$ are much smaller than $\Omega_1$ and $\Omega_2$, we can neglect that extra term for the tight-binding model. Actually, this approximation summarizes the condition that the RF detuning has to be sufficiently small that they do not affect the original system dynamics, which is the case if the detuning is on the order of, or smaller than the unmodulated resonator bandwidth, in our case few MHz.

Finally, the same procedure described for two-dimensional crystals can be fully generalized to arbitrary high dimensions, as long as the largest detuning between any two RF driving signals remains much smaller than the coupling rates (i.e. the largest RF detuning is on the order or smaller than the spectral bandwidth of the unmodulated resonator).

$$H = \sum_{l_1}\sum_{l_2}\cdots\sum_{l_d}\sum_i \frac{\Omega_i}{2}(c^\dagger_{l_1,l_2,\ldots,l_i,\ldots,l_d} c_{l_1,l_2,\ldots,l_i+1,\ldots,l_d} + h.c.)$$

And if all RF power are equal, then

$$H = \sum_{<i,j>} \frac{\Omega}{2}(c_i^\dagger c_j + h.c.)$$

where $<i,j>$ sum over the neighbor lattice.

**Optical transmission spectrum of high-dimensional tight-binding electro-optic frequency combs.** We derive the form of transmission spectrum using input-output theory. The equation of motion for mode $a_j$ for a one-dimensional frequency crystal is

$$\dot{a}_j = \left(-i\omega_j - \frac{\kappa}{2}\right) a_j - i\Omega \cos\omega t\, (a_{j-1} + a_{j+1})$$

where $\kappa = \kappa_e + \kappa_i$ is the linewidth of the cavity and $\kappa_e, \kappa_i$ is the waveguide-cavity loss rate, cavity intrinsic loss rate, respectively. Going to a rotating frame $a_j \to a_j e^{-i\omega_j t}$ and applying rotating wave approximation:

$$\dot{a}_j = \left(-\frac{\kappa}{2}\right) a_j - i\frac{\Omega}{2}(a_{j-1} + a_{j+1})$$

This is the equation of motion for the mode of each lattice point of one-dimensional crystal, which means that each frequency modes are coupled with neighboring modes. When we excite the crystal at single specific frequency $\omega_L$ around $\omega_0$, the equation for the pump photon is given by

$$\dot{a}_0 = \left(-\frac{\kappa}{2}\right) a_0 - i\frac{\Omega}{2}(a_{-1} + a_1) - \sqrt{\kappa_e}\,\alpha_{in} e^{-i\Delta t}$$

where $\Delta = \omega_L - \omega_0$ is the laser detuning. All the other modes are occupied by the hopping from the pump field. In reciprocal space the equation of motion for $a_k$ is given by

$$\dot{a}_k = \left(-i\Omega \cos k - \frac{\kappa}{2}\right) a_k - \sqrt{\frac{\kappa_e}{N_t}}\,\alpha_{in} e^{-i\Delta t}$$

This equation means that since we pump a single point on the real space of crystal, we are pumping all the momentum $\mathbf{k}$ in reciprocal space. In the laser rotating frame we obtain the steady state amplitude of $a_k$

$$a_k = \sqrt{\frac{\kappa_e}{N_t}}\,\alpha_{in} \frac{1}{-i\Omega \cos k + i\Delta - \kappa/2}$$

Which means that when the cavity is under-coupled ($\kappa_e < \kappa_i$), the field inside the cavity $a_j = \frac{1}{\sqrt{N_t}}\sum_k e^{ikj} a_k$ consists of a sets of extremely under-coupled mode $a_k$ with an external coupling strength $\kappa_e/N_t$. The optical transmission is therefore given by

$$\left|\frac{a_{out}}{\alpha_{in}}\right|^2 = \frac{\left|\alpha_{in} + \sqrt{\kappa_e}\sum_j a_j e^{-i(\omega_j - \omega_L)t}\right|^2}{\alpha_{in}} \approx 1 + 2\mathbf{Re}(\sqrt{\kappa_e}\, a_0)/\alpha_{in} = 1 + 2\mathbf{Re}\left(\sqrt{\frac{\kappa_e}{N_t}}\sum_k a_k\right)/\alpha_{in}$$

We can now apply the approximation that in the output signal $\alpha_{in}$ is much larger than the field out of the cavity because of the extremely under-couple of each $a_k$ and drop the fast oscillating term, leading to the final output spectrum:

$$T(\Delta) = 1 + 2\frac{\kappa_e}{N_t}\text{Im}\left(\sum_k \frac{1}{(\Delta - \Omega\cos k) + i\frac{\kappa}{2}}\right)$$

The same model can be applied to high-dimensional crystals, where we first consider the case of two-dimensional frequency crystals. We start from the equation of motion derived from the Hamiltonian in rotating frame without pump:

$$\dot{a}_j = \left(-\frac{\kappa}{2}\right)a_j - i\frac{\Omega_1}{2}\left(a_{j-1}e^{-i\delta_1 t} + a_{j+1}e^{i\delta_1 t}\right) - i\frac{\Omega_2}{2}\left(a_{j-1}e^{-i\delta_2 t} + a_{j+1}e^{i\delta_2 t}\right)$$

and multiply $\frac{1}{\sqrt{N_t}}e^{il\cdot\delta t}$ to the above equation to get

$$\dot{c}_l = \left(il\cdot\delta - \frac{\kappa}{2}\right)c_l - i\frac{\Omega_1}{2}\left(c_{l_1+1,l_2} + c_{l_1-1,l_2}\right) - i\frac{\Omega_2}{2}\left(c_{l_1,l_2+1} + c_{l_1,l_2-1}\right)$$

And since $il\cdot\delta \ll \kappa, \Omega_1, \Omega_2$, we neglect this term in the equation. This equation describes the coupling between each lattice point. The effect of pump can be introduced by considering the input photon to this crystal is at frequency $\omega_L$ which correspond to the mode $c_{0,0}$ and all the other photons are generated by the hopping from these input photons. For $l = (0,0)$ we have

$$\dot{c}_{0,0} = \left(-\frac{\kappa}{2}\right)c_{0,0} - i\frac{\Omega_1}{2}\left(c_{1,0} + c_{-1,0}\right) - i\frac{\Omega_2}{2}\left(c_{0,1} + c_{0,-1}\right) - \sqrt{\kappa_e}\frac{1}{\sqrt{N_t}}\alpha_{in}e^{-i\Delta t}$$

Where $\Delta = \omega_L - \omega_0$. Then we go to the reciprocal space by

$$c_k = \frac{1}{\sqrt{N_t^2}}\sum_l e^{-ik\cdot l}c_l \qquad c_l = \frac{1}{\sqrt{N_t^2}}\sum_k e^{ik\cdot l}c_k$$

and get the equation of motion in reciprocal space

$$\dot{c}_k = \left(-i\Omega_1\cos k_1 - i\Omega_2\cos k_2 + i\Delta - \frac{\kappa}{2}\right)c_k - \sqrt{\frac{\kappa_e}{N_t^2}}\frac{1}{\sqrt{N_t}}\alpha_{in}$$

With the similar argument in one-dimensional case, we have the output

$$T(\Delta) = \left|\frac{a_{out}}{\alpha_{in}}\right|^2 = 1 + 2\text{Re}\left(\sqrt{\frac{\kappa_e}{N_t^2}}\sqrt{N_t}\sum_k c_k\right) = 1 + 2\frac{\kappa_e}{N_t^2}\text{Im}\left(\sum_k \frac{1}{(\Delta - \sum_i\Omega_i\cos k_i) + i\frac{\kappa}{2}}\right)$$

Finally, it is directly possible to obtain the general result for a frequency crystal with dimension $d$. The spectral mode is

$$c_l = c_{(l_1,l_2,\ldots,l_d)} = \frac{1}{\sqrt{N_t^{d-1}}}a_{l_1+l_2+\cdots+l_d}e^{i(l_1\delta_1+l_2\delta_2+\cdots+l_d\delta_d)t} = \frac{1}{\sqrt{N_t^{d-1}}}a_{\Sigma l_i}e^{il\cdot\delta t}$$

With $l = (l_1, l_2, \ldots, l_d)$ and $\delta = (\delta_1, \delta_2, \ldots, \delta_d)$. With reciprocal space defined as $c_k = \frac{1}{\sqrt{N_t^d}}\sum_l e^{-ik\cdot l}c_l$ we obtain the equation of motion

$$\dot{c}_k = \left(-i\sum_i\Omega_i\cos k_i + i\Delta - \frac{\kappa}{2}\right)c_k - \sqrt{\frac{\kappa_e}{N_t^d}}\frac{1}{\sqrt{N_t^{d-1}}}\alpha_{in}$$

And transmission is

$$T(\Delta) = 1 + 2\frac{\kappa_e}{N_t^d}\text{Im}\left(\sum_k \frac{1}{(\Delta - \sum_{i=1}^d \Omega_i \cos k_i) + i\frac{\kappa}{2}}\right)$$

This consideration shows that for all frequency crystals, the basic picture is that the tight-binding coupling leads to a set of eigenstates $c_k$ and all of these $c_k$ are extremely under-coupled with a external coupling rate $\kappa_e/N_t^d$. Each of them contributes a small peak in transmission spectrum and the total transmission is a summation of these eigenmodes in reciprocal space.

At the same time, the DOS for tight-binding model given by the Green function, which is:

$$D(E) = -\frac{1}{\pi}\text{Im}\lim_{s\to 0^+}\sum_k \frac{1}{(E - \sum_i J_i \cos k_i) + is}$$

Comparing the transmissions spectrum and the DOS of the tight-binding model shows that the transmission has the same form of density of state under our high-Q cavity approximation (and we choose $\hbar = 1$):

$$T(\Delta) = 1 - 2\pi\frac{\kappa_e}{N_t^d}D(\Delta)$$